\documentclass[sigconf, nonacm]{acmart}
\usepackage{preprint}
\usepackage{xurl}
\usepackage{colortbl}
\hypersetup{hidelinks}
\newcommand{\Best}[1]{\begingroup\setlength{\fboxsep}{0pt}\colorbox{green!15}{\textbf{#1}}\endgroup}
\newcommand{\Second}[1]{\underline{#1}}
\DeclareRobustCommand{\authorrevision}[1]{#1}
\setcopyright{none}
\acmDOI{}
\acmISBN{}
\acmConference[]{}{}{}
\acmBooktitle{}
\authorsaddresses{}

\begin{document}
\title[ConTra: A Training Framework for Effective Contextual Retrieval via Prefix Generation]{ConTra: {A Training Framework for Effective Contextual Retrieval via Prefix Generation}}

%%
%% The "author" command and its associated commands are used to define the authors and their affiliations.
\author{Jiajun Hou}
\affiliation{%
  \institution{CUHK-Shenzhen}
  \city{Shenzhen}
  \country{China}
}
\email{jiajunhou@link.cuhk.edu.cn}

\author{Jingbang Chen}
\authornote{Corresponding author.}
\affiliation{%
  \institution{CUHK-Shenzhen \& SLAI}
  \city{Shenzhen}
  \country{China}
}
\email{chenjb@cuhk.edu.cn}

\begin{abstract}
Modern data management systems store and organize large collections of long documents, such as web pages, encyclopedia articles, and technical manuals, whose effective management is critical to the accuracy and reliability of downstream tasks including information retrieval, question answering, and retrieval-augmented generation. To make these document collections searchable at scale, systems typically split each document into smaller chunks and index them as independent records. However, chunking separates passages from the document-level context needed to identify and rank them. Existing methods address this missing context along two lines. Contextual encoders pack document context into dense chunk vectors, which leaves that context inaccessible to other downstream components. Prompting large language models to prepend a text prefix makes the context readable to both vector search and reranking, yet running large models over every chunk incurs high indexing overhead, and the generated prefixes often underperform specialized contextual encoders. To capture the cross-stage value of readable prefixes efficiently, we introduce \textsc{ConTra} (\emph{Context That Travels}), a reinforcement learning framework that trains lightweight language models to generate one concise prefix per chunk during index construction. Its reward combines feedback from frozen embedding and reranking models with factual verification and length control, and storing the generated prefix allows it to be reused across multiple stages. To evaluate contextual retrieval across workloads and interference levels, we introduce {RecoBench} (\emph{{Retrieval in Natural and Perturbed Contexts Benchmark}}), comprising five real-world collections and three controlled conditions that isolate distracting text. Across ConTEB and {RecoBench}, \textsc{ConTra}-2B outperforms open contextual encoders such as PPLX-4B, commercial models such as Voyage-4, and the prompted 9B baseline using Anthropic's method. {On ConTEB, \textsc{ConTra}-2B surpasses the best overall scores of the prompted 9B baseline and PPLX-4B by up to 3.15 points while indexing short documents $7.74\times$ and $2.15\times$ as fast as those baselines, respectively.} Post-trained only on ConTEB, it still achieves the highest overall reranking score on {RecoBench} and exhibits lower sensitivity to random and topically similar distractors.
Code, datasets, and trained models are available at \url{https://sstan1.github.io/ConTra-Context-That-Travels/}.
\end{abstract}
\maketitle
\pagestyle{plain}
\thispagestyle{plain}

\section{Introduction}
\label{sec:introduction}

Data management systems organize text collections spanning millions of source documents~\cite{DBLP:journals/pvldb/LiZXCWSFH26}, whose effective management is critical to the accuracy and reliability of downstream tasks such as finding relevant studies~\cite{DBLP:journals/pvldb/PatelJPGAGZ25}, extracting clauses from contracts~\cite{DBLP:journals/pvldb/ShankarCSPW25}, and generating literature digests~\cite{DBLP:journals/pvldb/PatelJPGAGZ25}.
To make long documents searchable at scale, {systems often split} them into smaller units, called \emph{chunks}~\cite{DBLP:journals/pvldb/LiZXCWSFH26,DBLP:conf/emnlp/Chen0C0MZ0024}, that can be stored as separate records and indexed for retrieval~\cite{DBLP:journals/pvldb/LiSPWXLN25,DBLP:journals/pvldb/SunCDJGHYWC25}.
{Retrieval-augmented generation (RAG) systems are commonly used to retrieve chunks from the index}~\cite{DBLP:conf/nips/LewisPPPKGKLYR020,DBLP:journals/pvldb/ZhouSSWWHZLLMF25}{. The retrieved chunks can be returned directly as search results or used as evidence to} answer questions~\cite{DBLP:journals/pvldb/WangZF26} and check factual claims~\cite{DBLP:conf/naacl/PetroniPFLYCTJK21,DBLP:journals/pvldb/PatelJPGAGZ25}.
{For RAG over large collections, the index must retrieve the evidence needed for an accurate answer}~\cite{DBLP:journals/pvldb/WangZF26,DBLP:journals/pvldb/ZhouSSWWHZLLMF25} {while remaining affordable to build and fast to query}~\cite{DBLP:journals/pvldb/LiZXCWSFH26}.
{In a \emph{multi-stage retrieval} pipeline, a system first uses} a vector index to shortlist chunks by embedding similarity~\cite{DBLP:journals/pvldb/LiSPWXLN25,DBLP:journals/pvldb/LiZXCWSFH26}. {To improve ranking accuracy, the system often uses a reranker to score} each shortlisted chunk's text {for relevance to the query and reorder the candidates by that score}~\cite{DBLP:journals/pvldb/LiZXCWSFH26,DBLP:conf/sigir/MacdonaldFPM25}.
{However, conventional chunking methods can separate a passage from the context that identifies it}~\cite{DBLP:journals/pvldb/WangZF26}. For example, consider a query for a company's operating income in a particular fiscal year~\cite{DBLP:journals/corr/abs-2311-11944}{: the financial table containing the figure may rely on a preceding heading to identify the company and year.}
{If that heading falls in another chunk, vector search may retrieve a similar table for the wrong company or period, and a reranker reading the table alone lacks the evidence to correct the error.}
To improve retrieval quality at scale, systems must address the fundamental data management challenge of keeping chunks independently searchable while preserving the document context needed to find the right candidates and rank them correctly.

\begin{figure*}[t]
\centering
\includegraphics[width=\textwidth]{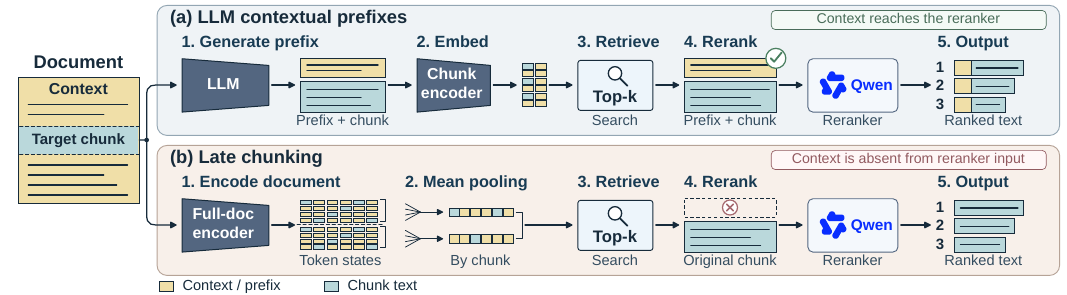}
\caption{A textual prefix carries document context through embedding retrieval and reranking. Contextual encoders enrich chunk vectors, while the reranker receives only the original chunk text.}
\Description{The upper pipeline supplies a generated prefix and the original chunk to both the embedding model and reranker. The lower pipeline contextualizes vectors but supplies original chunks to the reranker.}
\label{fig:context-travels}
\end{figure*}

Such a challenge of preserving document context has attracted considerable attention from data scientists~\cite{DBLP:conf/emnlp/ContiFVBHC25,DBLP:journals/corr/abs-2409-04701,DBLP:journals/pvldb/WangZF26} as well as from AI companies such as Perplexity~\cite{DBLP:conf/acl/EslamiGKMWB26}, Voyage AI~\cite{voyage2025context3}, and Anthropic~\cite{anthropic2024contextual}.
Two lines of work address this challenge by encoding document context into chunk vectors or preserving it explicitly in text.
Figure~\ref{fig:context-travels} illustrates how these approaches differ in the context available to the vector index and the reranker.
The first line of work incorporates surrounding document information into chunk vectors via late chunking~\cite{DBLP:journals/corr/abs-2409-04701}.
Building on this design, InSeNT~\cite{DBLP:conf/emnlp/ContiFVBHC25} trains encoders to distinguish chunks from the same document, and PPLX~\cite{DBLP:conf/acl/EslamiGKMWB26} adapts an autoregressive large language model for contextual encoding.
{These methods can improve candidate selection by changing only the vector representation of each chunk.}
However, this line of work faces several limitations.
PPLX, for example, requires substantial additional pretraining to enable bidirectional attention, adding to the time and computation required to train the encoder.
Furthermore, a reranker that reads the original chunk text cannot use document context encoded only in the chunk vector.
The second line of work preserves document context explicitly in text.
Anthropic's method prompts a large language model to read the source document and generate a prefix for each chunk, supplying the augmented text to both retrieval and reranking~\cite{anthropic2024cookbook}.
The prefixes can be reused across queries, a benefit of enriching the index offline~\cite{DBLP:journals/corr/abs-2504-03598}.
The upfront cost is the substantial time and computation needed to generate a prefix for every chunk during index construction. {Prompted prefixes have also underperformed leading contextual encoders in prior retrieval evaluations~\cite{DBLP:conf/acl/EslamiGKMWB26}.}

{However, those evaluations primarily examined single-stage retrieval, where chunks are ranked by embedding similarity alone, overlooking multi-stage retrieval. We argue that because an LLM-generated prefix makes context available as readable text beyond the chunk vector, it can be reused across retrieval stages and may offer greater benefits than embedding-only evaluations suggest. In Section~\ref{sec:observation}, we use controlled experiments to measure the contribution of prefixes at each stage and demonstrate this potential.}

{To realize this potential at a lower indexing cost, we introduce} \textsc{ConTra} (\emph{Context That Travels}), a novel reinforcement learning framework for training lightweight language models to write concise, chunk-specific prefixes for both candidate selection and reranking (Section~\ref{sec:method}).
At indexing time, the trained model generates one prefix per chunk.
\textsc{ConTra} prepends the prefix to the original text and stores the augmented text and its embedding under the original chunk identifier.
At query time, the vector index uses the embedding to select candidates, and the reranker reads the same augmented text to refine their order.
During training, frozen embedding and reranking models evaluate each sampled prefix by scoring the augmented chunk against the training query.
A {frozen judge model} checks {whether the} source {supports the information added by each prefix,} and {a length penalty limits prefix length and reduces competition among chunks from} the {same document}.
{Together, these signals let small models generate accurate prefixes that distinguish the target chunk.}
{To accommodate different indexing budgets across deployments, we train} \textsc{ConTra}-0.8B, \textsc{ConTra}-2B, and \textsc{ConTra}-4B.

Existing benchmarks such as ConTEB~\cite{DBLP:conf/emnlp/ContiFVBHC25} emphasize cases where document context is needed to identify relevant chunks.
In practice, document collections mix self-contained and context-dependent chunks, and surrounding text can introduce competing entities or topics.
Evaluating contextual retrieval therefore requires testing both how it uses helpful context and how it handles distracting text.
To this end, we introduce {RecoBench} (Section~\ref{sec:real-data}), {a contextual chunk retrieval benchmark spanning web search, technical QA, story comprehension, meetings, and financial analysis{, with human annotations mapped to relevant chunks. A separate controlled evaluation holds queries, target chunks, and relevance labels fixed while presenting each chunk in isolation, alongside randomly sampled passages, or with LLM-generated, topically similar but unrelated passages.}}

Our contributions are summarized as follows:

\noindent\textbf{(1) ConTra Framework:} {Our controlled experiments demonstrate that readable prefixes can be effectively reused across candidate retrieval and reranking, revealing their potential for contextual retrieval.} {Building on this finding, we} propose \textsc{ConTra}, a reinforcement learning framework combining online embedding and reranking evaluation, factual verification, and length control to train prefix generators across multiple model scales. By encouraging informative, source-supported, and concise prefixes, \textsc{ConTra} improves {contextual retrieval quality} while lowering indexing cost.

\noindent\textbf{(2) RecoBench Benchmark:} We introduce {RecoBench}, {a benchmark for evaluating contextual retrieval on {real-world document collections} and under {controlled} interference. Whereas ConTEB emphasizes context-dependent cases, our five application collections retain queries with varying context needs. Three controlled conditions hold queries and target chunks fixed to measure the effect of adding random or topically similar distractors. {\authorrevision{RecoBench fills a gap in existing evaluations by covering real-world settings and noisy datasets, both essential for testing contextual retrieval models and advancing their practical use.}}}

\noindent\textbf{(3) Extensive Experiments:} {Through extensive experiments, we show that} \textsc{ConTra} delivers state-of-the-art end-to-end retrieval quality with substantially lower indexing cost and that {RecoBench} reveals which methods remain effective across workloads and under interference.
{Trained only on ConTEB, \textsc{ConTra}-2B outperforms the open PPLX-4B encoder, commercial Voyage-4, and the prompted 9B baseline using Anthropic's method on both benchmarks. It achieves the highest overall reranking score on {RecoBench} and surpasses the best overall ConTEB scores of the 9B and PPLX baselines by up to 3.15 points.} \textsc{ConTra}-2B also indexes short documents $7.74\times$ as fast as the 9B pipeline and $2.15\times$ as fast as PPLX-4B, while post-training with reinforcement learning reduces total indexing time by up to 52.3\% at a fixed 2B model scale.\looseness=-1

\noindent\textbf{Outline.} Section~\ref{sec:introduction} motivates the study with context loss from chunking and summarizes our contributions. Section~\ref{sec:related} reviews retrieval pipelines and contextual retrieval. Section~\ref{sec:observation} isolates prefix effects on retrieval and reranking. Section~\ref{sec:method} introduces \textsc{ConTra}'s reinforcement learning framework and reward design. Sections~\ref{sec:experiments} and~\ref{sec:real-data} report experimental results and ablations, and introduce our proposed RecoBench benchmark along with its evaluation results. Sections~\ref{sec:efficiency} and~\ref{sec:cases} cover indexing costs, storage, and case studies. Section~\ref{sec:conclusion} concludes.\looseness=-1

\section{Related Work}
\label{sec:related}
\subsection[Retrieval Models and Pipelines]{{Retrieval Models and Pipelines}}
{Evaluating complex relevance models across an entire large corpus is too costly at query time. Modern retrieval pipelines, as supported by frameworks such as PyTerrier-RAG and RAGPerf~\cite{DBLP:journals/pvldb/LiZXCWSFH26,DBLP:conf/sigir/MacdonaldFPM25}, therefore combine fast candidate retrieval with compute-intensive reranking to balance efficiency and accuracy~\cite{DBLP:conf/sigir/KhattabZ20}. Given a query, a retriever first selects a small candidate set using an embedding model, often paired with lexical matching in multi-way retrieval~\cite{DBLP:conf/sigir/FormalPC21,DBLP:conf/iclr/XiongXLTLBAO21}. While BM25 relies on term overlap and corpus statistics to match exact names and identifiers~\cite{DBLP:journals/ftir/RobertsonZ09}, queries and passages frequently express the same meaning with different words.}

{Embedding models address this lexical gap by mapping queries and passages into a shared continuous vector space to capture semantic similarity~\cite{DBLP:conf/emnlp/KarpukhinOMLWEC20,DBLP:conf/iclr/XiongXLTLBAO21,DBLP:journals/tmlr/IzacardCHRBJG22,DBLP:conf/emnlp/Ni0LDAMZLHCY22}. Early architectures build upon bidirectional BERT encoders~\cite{DBLP:conf/naacl/DevlinCLT19,DBLP:journals/tacl/LuanETC21,DBLP:conf/emnlp/ReimersG19}, whereas recent models such as E5-Mistral and Qwen3-Embedding {draw on the knowledge encoded in pretrained LLMs}~\cite{DBLP:conf/acl/WangYHYMW24,DBLP:journals/corr/abs-2506-05176}.}\looseness=-1

{Because these retrievers only narrow the corpus down to a candidate set, a reranker must then judge fine-grained relevance by jointly evaluating the query and candidate text~\cite{DBLP:conf/sigir/Zhuang0J0MLNWB23,DBLP:journals/corr/abs-2312-02724}. Cross-encoders such as BERT rerankers and Qwen3-Reranker evaluate individual query--passage pairs independently~\cite{DBLP:journals/corr/abs-1901-04085,DBLP:journals/corr/abs-2506-05176}, while monoT5 formulates this judgment as relevance-token generation~\cite{DBLP:conf/emnlp/NogueiraJPL20}. Rather than judging pairs in isolation, RankGPT compares a group of candidates through listwise ordering~\cite{DBLP:conf/emnlp/0001YMWRCYR23}, whereas RankRAG unifies reranking with answer generation~\cite{DBLP:conf/nips/YuP0WYZSC24}. {The retriever selects the candidates, and the reranker determines their final order.}}\looseness=-1

\subsection[Contextual Retrieval]{{Contextual Retrieval}}
{Contextual retrieval mitigates the information loss caused by partitioning documents into isolated chunks. The first paradigm encodes surrounding document context directly into chunk vectors. Late chunking achieves this by encoding an entire source window before pooling token representations within each chunk boundary~\cite{DBLP:journals/corr/abs-2409-04701}. Extending this approach, InSeNT trains contextual encoders using in-document negatives, encouraging models to preserve passage-specific details alongside shared document context~\cite{DBLP:conf/emnlp/ContiFVBHC25}. Adapting this paradigm to autoregressive LLMs requires overcoming causal attention, which restricts tokens to preceding context. PPLX adapts an LLM backbone via diffusion pretraining on roughly 250 billion tokens to enable bidirectional attention, followed by contrastive retrieval training~\cite{DBLP:conf/acl/EslamiGKMWB26}. Similarly, Voyage-context-3 and Voyage-context-4 provide document-aware chunk embeddings via proprietary APIs~\cite{voyage2025context3,voyage2026context4}. However, PPLX incurs substantial additional pretraining overhead, and storing context implicitly in vectors leaves downstream components reading raw text---such as rerankers---without access to that context.}

{The second paradigm preserves context explicitly as readable text. Anthropic prompts an LLM to generate a concise prefix situating each chunk within its source document, prepending it to the passage for indexing and reranking~\cite{anthropic2024contextual}. Because prefixes are materialized at indexing time, they are reusable {for retrieval and reranking across queries} while remaining directly legible to rerankers. The same prefix can therefore be reused across retrieval stages and backbones, independent of the embedding space in which a chunk is indexed. Nevertheless, prompting large LLMs incurs high indexing latency and computational cost, whereas smaller prompted models often produce verbose or imprecise prefixes that limit retrieval gains. \textsc{ConTra} addresses both bottlenecks by training prefix generators with reinforcement learning, combining {feedback from the embedding model and reranker}, factual verification, and length control to produce concise, accurate prefixes that boost retrieval accuracy while reducing indexing time.}

\section[Observation: The Role of Prefixes in Contextual Retrieval]{Observation: The Role of Prefixes in {Contextual} Retrieval}
\label{sec:observation}

{We expect retrieval to benefit when several stages can reuse the same readable prefix. Anthropic tried this approach in its contextual retrieval pipeline~\cite{anthropic2024contextual} but did not investigate it in depth. On ConTEB, we vary which stages receive the prefix to measure its contribution at each stage (Table~\ref{tab:motivation}). Our analysis shows that a readable prefix can serve multiple retrieval stages, with complementary gains. Even with prefixes generated by an untrained 2B model, the resulting pipeline gives better final rankings than retrieving with PPLX-4B and reranking raw chunks.}

\begin{table}[!t]
\centering
\caption{{Controlled comparison of prefix placement on ConTEB. {Emb. and Rer. report nDCG@10 before and after reranking, respectively. Rec.@15 is the fraction of all chunks labeled relevant to the query that appear among the 15 candidates.}}}
\label{tab:motivation}
\begingroup\fontsize{9.5}{10.4}\selectfont
\setlength{\tabcolsep}{2.5pt}\renewcommand{\arraystretch}{1.0}
\begin{tabular*}{\columnwidth}{@{\extracolsep{\fill}}llccc@{}}
\toprule
\multicolumn{1}{c}{Retrieval view} & \multicolumn{1}{c}{Reranker view} & Emb. & Rec.@15 & Rer. \\
\midrule
Raw & Raw & .5484 & .7691 & .6483 \\
Prefix + raw & Raw & .6547 & .8475 & .6924 \\
Raw & Prefix + raw & .5484 & .7691 & .6940 \\
Prefix + raw & Prefix + raw & .6547 & .8475 & .7585 \\
\midrule
PPLX-4B & Raw & .8210 & .9387 & .7310 \\
PPLX-4B & Prefix + raw & .8210 & .9387 & .8135 \\
\bottomrule
\end{tabular*}\endgroup
\end{table}

\noindent\textbf{{Experimental setup for comparing prefix placement.}}
{\emph{Raw} denotes the original chunk, and \emph{Prefix + raw} prepends a prefix generated by an untrained Qwen3.5-2B model using Anthropic’s prompt. Qwen3-Embedding-4B retrieves 15 candidates and Qwen3-Reranker-4B scores them. {We give the prefix to the embedding model alone, the reranker alone, both models, or neither.} PPLX-4B retrieves its candidates using native contextual embeddings. We hold the candidates fixed when varying the reranker input.}

\noindent\textbf{{Prefixes improve embedding retrieval and supply more gold chunks to the reranker.}}
{Prepending document context before encoding helps the embedding model distinguish a relevant chunk from similar passages, strengthening initial retrieval ranking. This better ranking brings more gold chunks into the reranker's candidate pool, raising Recall@15 from 0.7691 to 0.8475. Even when the reranker reads only raw chunk text, the stronger candidate pool raises final nDCG@10 from 0.6483 to 0.6924.}

\noindent\textbf{Prefixes directly {help the reranker judge relevance}.}
{Even without bringing additional gold chunks into the shortlist, prefixes improve the reranker's relevance judgments. We rerank the same candidates retrieved from raw chunks twice, changing only whether the reranker sees the prefix. Making the prefix visible raises final nDCG@10 from 0.6483 to 0.6940, showing that the added context helps the reranker recognize and promote relevant chunks already in the shortlist.}

\noindent\textbf{{Higher candidate recall alone does not guarantee a better final ranking.}}
{PPLX-4B achieves strong initial retrieval at 0.9387 Recall@15 and 0.8210 embedding nDCG@10, bringing more gold chunks into the candidate pool than retrieving with prefixes from an untrained 2B model. Yet when the reranker reads only raw text, the PPLX-4B pipeline yields only 0.7310 final nDCG@10, trailing the 0.7585 achieved when both Qwen3-Embedding-4B and Qwen3-Reranker-4B read those prefixes. PPLX-4B encodes document context in the chunk vectors, so the information that helped select a passage is unavailable to a reranker reading the passage in isolation. Attaching the same readable prefixes to the fixed PPLX-4B candidates tests this explanation. Without retrieving a single additional candidate, the reranker can then rank the relevant chunks higher, raising final nDCG@10 to 0.8135.}

\noindent\textbf{{Readable prefixes can be reused across retrieval stages.}}
{Applying the same readable prefix at either stage improves final nDCG@10 from 0.6483 to roughly 0.69. Reusing it across embedding retrieval and reranking reaches 0.7585, a further 6.45-point gain over the stronger of the two single-placement settings. These complementary gains allow even an untrained 2B prefix generator to surpass PPLX-4B retrieval with raw-text reranking, demonstrating the potential of readable prefixes in multi-stage retrieval.}

\section{\textsc{ConTra}: Learning to Write Context Prefixes}
\label{sec:method}
{To realize the potential of LLMs in multi-stage retrieval, we introduce \textsc{ConTra}, a framework that trains them to generate more effective context prefixes.} For each training instance, the generator reads a source window and target chunk, samples candidate prefixes, and prepends each candidate to the original chunk. Frozen embedding and reranking models then evaluate how well the augmented text matches a training query. A factual judge verifies whether the added information is supported by the source and correctly attributed to the target, while length control discourages superfluous details. We combine these signals into a reward to update the generator. After training, the model generates prefixes during indexing, and the resulting text is reused for embedding retrieval and reranking across queries.

\subsection{Indexing and Retrieval with Context Prefixes}
\label{sec:workflow}
Let $D_i$ denote a source document and $c_i$ denote a target chunk. Our generator $\pi_\theta$ receives $x_i = (W_i, c_i)$, where $W_i$ contains the full document when it fits the source budget, or a window of adjacent complete chunks in source order. The prompt marks the target within this window and also presents it separately, allowing the model to relate the target to a preceding heading or {speaker's earlier remark}. Conditioned on this input, the model generates a prefix $p_i$, which is concatenated with the original chunk:
\begin{equation}
\begin{aligned}
p_i&\sim\pi_\theta(\,\cdot\mid x_i),\\
\widetilde c_i&=p_i\mathbin{\Vert}c_i.
\end{aligned}
\label{eq:prefix-generation}
\end{equation}
where $\parallel$ denotes concatenation. Because $p_i$ depends exclusively on $(W_i, c_i)$ without query conditioning, each stored prefix is reusable across arbitrary incoming queries. The original text preserves verbatim passage evidence, while the prefix supplies missing identities or relationships.

At indexing time, we generate one prefix per chunk and store both the augmented text and its vector $\mathbf e_i = f(\widetilde c_i)$ under the original chunk identifier, where $f$ is an $\ell_2$-normalized embedding model and $z$ is the reranker's relevance logit. Given a corpus of $N$ chunks and an incoming query $q$, the pipeline computes:
\begin{equation}
\begin{aligned}
\mathcal C_K(q)
 &=\operatorname{TopK}_{i\in\{1,\ldots,N\}}
       \cos\!\left(f(q),\mathbf e_i\right),\\
\widehat{\mathcal C}_K(q)
 &=\operatorname{argsort}^{\downarrow}_{i\in\mathcal C_K(q)}
       z(q,\widetilde c_i).
\end{aligned}
\label{eq:retrieval-pipeline}
\end{equation}
{The vector index} retrieves $K$ identifiers by vector similarity. Each identifier links the retrieved vector to its augmented text, enabling the reranker to read the prefix and original passage jointly before ordering candidates by relevance.

\subsection{Training Data}
\label{sec:task}
We construct our training pool from the ConTEB training split~\cite{DBLP:conf/emnlp/ContiFVBHC25}, where each triple $(D_i, c_i, q_i)$ pairs a source document $D_i$ and target chunk $c_i$ with a relevant query $q_i$ for reward scoring. During training, a generator can easily converge toward broad, generic descriptions: although low-risk and easy to produce, high-level summaries offer limited retrieval gains and fail to distinguish closely related chunks. Conversely, specific identifying details---such as exact entities, dates, or cross-chunk relations---naturally situate a passage within its broader context while supplying the discriminative cues required for accurate retrieval. To ensure the training data rewards specific details over generic descriptions, we curate query--target pairs via source-based proposals and two screening checks.

Starting from 100,000 query--target pairs randomly sampled from the ConTEB training set, GPT-6 Luna~\cite{openai2026luna} reads $(D_i, c_i)$, proposes a missing entity, date, or relation needed to interpret the target, and locates its supporting source span. For each supported proposal, it writes both a specific prefix $p_i^{\mathrm{spec}}$ containing that identifying detail and a generic description $p_i^{\mathrm{gen}}$ summarizing the broader topic. We then apply two screening checks to retain instances with clear RL learning signals and a measurable retrieval advantage from specific details.

First, inspired by DAPO's filtering of zero-variance reward groups~\cite{DBLP:conf/nips/YuZZYZYDFLLLLLL25}, we verify that the policy can explore meaningful contextual variations: a separate 2B proposal model generates eight candidate prefixes per input. Let $\mathcal{V}_i$ contain the indices of valid responses and $\mathcal{S}_i = (S(p_{ij}))_{j \in \mathcal{V}_i}$ denote their screening reward sequence. We retain instances satisfying:
\begin{equation}
\begin{aligned}
Q_{0.9}(\mathcal S_i)-Q_{0.1}(\mathcal S_i)&\geq0.05,\\
|\mathcal V_i|\geq6,\qquad
|\{p_{ij}:j\in\mathcal V_i\}|&\geq2.
\end{aligned}
\label{eq:data-variation}
\end{equation}
where $Q_\alpha$ denotes the $\alpha$-quantile. The quantile gap ensures sufficient reward variation across usable alternatives without allowing a single outlier to dominate. The screening reward combines retrieval gains, source support, and prefix length, where the factual term compares the verification rewards of the augmented and original target against the same document.

Second, we verify that the specific detail measurably improves retrieval over generic context. Let $n_i(p)$ denote nDCG@10 with the target as the sole relevant chunk for $q_i$. We retrieve and rerank 15 candidates in the training corpus for each text view, changing only the target's prefix while keeping competing chunks unchanged. We retain instances satisfying:
\begin{equation}
\begin{aligned}
n_i(p_i^{\mathrm{spec}})-n_i(p_i^{\mathrm{gen}})&\geq0.05,\\
n_i(p_i^{\mathrm{spec}})&\geq n_i(\varnothing),
\end{aligned}
\label{eq:data-specificity}
\end{equation}
where $\emptyset$ leaves the target unchanged. The first condition ensures that specific identifying details substantially outperform generic summaries, while the second requires them to match or improve upon the raw chunk.

Together, these two screening checks filter the 100,000 randomly sampled candidates down to 5,760 query--target pairs. Each \textsc{ConTra} model is initialized from Qwen3.5 and trained via RL for one epoch.

\subsection{Reward Design}
\label{sec:reward}
{Our reward} combines embedding feedback, which aligns the target chunk with the query, with reranking feedback, which refines fine-grained relevance distinctions. Because relevance scores alone can favor unsupported associations or verbose, generic summaries, we pair {the two relevance signals} with factual grounding and length control:
\begin{equation}
\begin{aligned}
R(p_i;D_i,c_i,q)={}&\underbrace{R_{\mathrm{emb}}+R_{\mathrm{rank}}}_{\text{{retrieval utility}}}\\
&+\underbrace{R_{\mathrm{fact}}}_{\text{factual grounding}}
 +\underbrace{R_{\mathrm{len}}}_{\text{length control}}.
\end{aligned}
\label{eq:reward}
\end{equation}
The generator must output a complete plain-text prefix for direct concatenation. Malformed or truncated outputs receive a fixed penalty of $-1.5$---strictly below the theoretical minimum of $-1.475$ for valid prefixes---making structural validity a prerequisite for earning retrieval utility.

\subsubsection[Embedding and Reranking Rewards]{{Embedding and Reranking Rewards}}
\label{sec:utility-module}
Both utility branches compare the augmented chunk $\widetilde c_i$ against the raw chunk $c_i$ to isolate the marginal contribution of the generated prefix.

For the embedding branch, training queries often contain implicit pronouns or references, such as ``he'' or ``the facility,'' that rely on document context to resolve. Scoring prefixes solely against the raw query embedding $f(q)$ can under-reward additions that supply the missing entity identity, whereas relying solely on a context-wrapped query risks shifting the target vector away from the specific question toward general document-opening semantics. We therefore encode the raw query $q$ and a context-wrapped query $\mathrm{ctx}(q, D_i)$ separately into embeddings of identical dimension and take their normalized weighted sum:
\begin{equation}
\mathbf q^*=\operatorname{norm}\bigl(0.75f(q)+
                         0.25f(\operatorname{ctx}(q,D_i))\bigr),
\label{eq:query}
\end{equation}
where $\mathrm{ctx}(q, D_i)$ brackets the opening of the document's first chunk, up to 128 words, with the query on both ends under a fixed disambiguation instruction, and $\mathrm{norm}(\cdot)$ applies $\ell_2$ normalization. Retaining a 0.75 weight preserves the core question intent, while the 0.25 contextual embedding supplies disambiguating entity cues. Substituting $\mathbf q^*$ for $f(q)$, we compute the embedding reward as:
\begin{equation}
R_{\mathrm{emb}}=\operatorname{clip}\!\left(
\cos(\mathbf q^*,f(\widetilde c_i))-
\cos(\mathbf q^*,f(c_i)),\,-1,\,1\right).
\label{eq:embedding-reward}
\end{equation}
Subtracting raw chunk similarity establishes a per-instance baseline, rewarding prefixes that pull the target closer to the resolved query. Evaluating this change directly on the target chunk keeps online RL training tractable without full-corpus retrieval, which is reserved for the offline check in Equation~\ref{eq:data-specificity} and final evaluation, whereas query-time inference encodes raw $f(q)$ as in Equation~\ref{eq:retrieval-pipeline}.

While embedding similarity captures coarse topical alignment, it can miss fine-grained relational distinctions, such as when two prefixes match the same topic but attribute a statement to different entities. Because the reranker jointly attends to the query and the augmented chunk, we add a reranking utility branch derived from its native relevance logit difference:
\begin{equation}
R_{\mathrm{rank}}=0.025\tanh\!\left(
                       z(q,\widetilde c_i)-z(q,c_i)\right).
\label{eq:rerank-reward}
\end{equation}
Raw reranker logits are unbounded and highly sensitive to minor wording edits; applying $\tanh$ preserves the direction of relevance gain while bounding extreme logit spikes within $(-1, 1)$. Scaling by 0.025 constrains $R_{\mathrm{rank}}$ within $[-0.025, 0.025]$, allowing it to act as a fine-grained tie-breaker among topically similar prefixes without overwhelming the smoother embedding signal.

\subsubsection{Factual Grounding}
\label{sec:grounding-module}
Relevance utility alone cannot distinguish a truthful prefix from a hallucinated one: inserting fabricated entities, dates, or roles that echo the query often increases both embedding similarity and reranker logits. For instance, if a query asks about a specific executive's decision, prepending that executive's name to a chunk describing someone else elevates both $\cos(\mathbf q^*, f(\widetilde c_i))$ and $z(q, \widetilde c_i)$ even though the claim contradicts the source.

To prevent such hallucinations, a frozen Qwen3.5-4B judge reads the target chunk $c_i$, generated prefix $p_i$, and up to 8,000 tokens of surrounding source evidence $\mathcal{E}_i$ to evaluate contextual faithfulness~\cite{DBLP:conf/eacl/ESJAS24,DBLP:conf/naacl/Saad-FalconKPZ24}, verifying both factual support in the source and accurate attribution to $c_i$. Its Yes/No logits define the support confidence:
\begin{equation}
s_i=\frac{\exp(a_{\mathrm{Yes}})}
         {\exp(a_{\mathrm{Yes}})+\exp(a_{\mathrm{No}})}
   =\sigma(a_{\mathrm{Yes}}-a_{\mathrm{No}}).
\label{eq:grounding-confidence}
\end{equation}
which we convert into a one-sided penalty:
\begin{equation}
R_{\mathrm{fact}}=-0.25\frac{\max(0.75-s_i,0)}{0.75}.
\label{eq:fact-reward}
\end{equation}
We design $R_{\mathrm{fact}}$ to vanish once $s_i \ge 0.75$: after a prefix achieves sufficient factual support, retrieval utility and length alone decide which identifying detail is most advantageous, whereas rewarding confidence above 0.75 would favor safe, generic truisms that are trivial to verify yet unhelpful for retrieval. Meanwhile, scaling the maximum penalty to $-0.25$ ensures that factual errors outweigh spurious utility gains; for example, at $s_i = 0.45$, $R_{\mathrm{fact}}$ equals $-0.10$, four times the maximum positive reranking reward.

\subsubsection{Length Control}
\label{sec:compactness-module}
Because utility is scored independently on each query--target pair, a generator can raise local similarity by collecting broad document-level details shared across neighboring chunks. When applied across an entire corpus, however, prepending every chunk with overlapping background context blurs distinctions between adjacent passages and causes them to compete for the same queries. Length control discourages this behavior by requiring selectivity:
\begin{equation}
R_{\mathrm{len}}=-0.2\min\!\left(
                      \frac{\max(\ell_i-16,0)}{96},1\right).
\label{eq:length-reward}
\end{equation}
where $\ell_i$ denotes the prefix length in generator tokens. Preliminary analysis showed that concise, informative prefixes average roughly 16 tokens, so leaving $\ell_i \le 16$ unpenalized accommodates essential identifying additions without friction. Beyond 16 tokens, the penalty increases linearly up to a cap of $-0.2$ at 112 tokens, ensuring that additional tokens are retained only when they contribute commensurate retrieval utility rather than generic filler. Keeping prefixes short also reduces autoregressive decoding steps during generation and lowers encoder token overhead at indexing time.

\subsection{Policy Optimization}
\label{sec:policy}
We optimize \textsc{ConTra} using DAPO~\cite{DBLP:conf/nips/YuZZYZYDFLLLLLL25}. For each training input, the policy samples a group of $G = 16$ candidate prefixes, which are scored by the frozen embedding model, reranker, and factual judge via Equation~\ref{eq:reward}. Each rollout batch comprises 32 inputs, totaling 512 sampled responses, and only the generator parameters are updated.

For prefix $j$ within sampling group $\mathcal{G}(j)$, its advantage is computed by centering its reward around the group mean~\cite{DBLP:journals/corr/abs-2402-03300}:
\begin{equation}
A_j=R_j-\frac{1}{G}\sum_{k\in\mathcal G(j)}R_k.
\label{eq:advantage}
\end{equation}
We omit group-wise standard deviation normalization so that the true magnitude of reward differences is preserved across completions. Following DAPO, sampling groups in which all candidates receive identical rewards---yielding zero policy gradient---are dynamically filtered and replaced during rollout. The policy is updated via a token-averaged clipped surrogate objective that reinforces above-average prefixes while suppressing below-average ones, using importance-sampling ratio clipping bounds of $[0.8, 1.28]$ and a dual-clipping threshold of $3.0$ for negative advantages.

% Declare the wide result table here so it floats beside the experimental section.
\begin{table*}[t]
\centering
\caption{ConTEB nDCG@10 for embedding retrieval (Emb.) and reranking (Rer.). Qwen3.5 rows {use} Anthropic's prompt. Green bold and underlining indicate the best and second-best scores, respectively.}
\label{tab:conteb-subsets}\label{tab:conteb-main}\label{tab:training-effect}
\begingroup
\fontsize{9.5}{10.4}\selectfont
\setlength{\tabcolsep}{1.3pt}
\renewcommand{\arraystretch}{1.13}
\begin{tabular*}{\textwidth}{@{\extracolsep{\fill}}l*{16}{c}cc@{}}
\toprule
Method & \multicolumn{2}{c}{\textbf{Ins.}} & \multicolumn{2}{c}{\textbf{COVID}} & \multicolumn{2}{c}{\textbf{Geo.}} & \multicolumn{2}{c}{\textbf{ESG}} & \multicolumn{2}{c}{\textbf{Football}} & \multicolumn{2}{c}{\textbf{SQuAD}} & \multicolumn{2}{c}{\textbf{NQA}} & \multicolumn{2}{c}{\textbf{MLDR}} & \multicolumn{2}{c}{\textbf{Overall}}\\
\cmidrule(lr){2-3}\cmidrule(lr){4-5}\cmidrule(lr){6-7}\cmidrule(lr){8-9}\cmidrule(lr){10-11}\cmidrule(lr){12-13}\cmidrule(lr){14-15}\cmidrule(lr){16-17}\cmidrule(lr){18-19}
 & Emb. & Rer. & Emb. & Rer. & Emb. & Rer. & Emb. & Rer. & Emb. & Rer. & Emb. & Rer. & Emb. & Rer. & Emb. & Rer. & Emb. & Rer.\\\midrule
Qwen3-Emb.-4B & .145 & .166 & .546 & .692 & .719 & .853 & .397 & .483 & .317 & .465 & .726 & .798 & .777 & .846 & .760 & .883 & .5484 & .6483\\
InSeNT & .451 & .394 & .360 & .538 & .831 & .885 & .230 & .466 & .252 & .367 & .684 & .790 & .667 & .793 & .719 & .872 & .5242 & .6381\\
PPLX-0.6B & \Second{.997} & .333 & \Second{.561} & \Second{.700} & .898 & .880 & .461 & .541 & .714 & .650 & .789 & .838 & \Second{.842} & .861 & .867 & .885 & .7662 & .7110\\
PPLX-4B & \Best{1.000} & .344 & \Best{.620} & \Best{.741} & .935 & .879 & \Best{.631} & .602 & .781 & .677 & \Best{.837} & \Second{.850} & \Best{.869} & \Best{.865} & \Second{.895} & .890 & \Best{.8210} & .7310\\
Voyage-3 & \Best{1.000} & .348 & .547 & .695 & .928 & .880 & .574 & .571 & \Second{.796} & .728 & .826 & \Second{.850} & .818 & .861 & \Best{.896} & .900 & \Second{.7983} & .7292\\
Voyage-4 & .994 & .344 & .554 & .689 & \Best{.938} & .884 & \Second{.579} & .581 & \Best{.802} & .727 & \Second{.830} & \Best{.853} & .816 & \Second{.862} & .873 & .901 & .7981 & .7300\\
\midrule
Qwen3.5-0.8B & .154 & .199 & .487 & .664 & .746 & .859 & .427 & .498 & .310 & .460 & .719 & .785 & .781 & .843 & .807 & .916 & .5538 & .6531\\
Qwen3.5-2B & .708 & \Second{.809} & .526 & .680 & .798 & .886 & .449 & .607 & .405 & .538 & .735 & .774 & .779 & .843 & .837 & .930 & .6547 & .7585\\
Qwen3.5-4B & \Second{.997} & \Best{1.000} & .523 & .680 & .925 & .928 & .543 & .664 & .601 & .742 & .669 & .719 & .764 & .834 & .797 & .921 & .7274 & .8110\\
Qwen3.5-9B & .994 & \Best{1.000} & .547 & .691 & .910 & .945 & .552 & .668 & .543 & .700 & .743 & .830 & .778 & .841 & .835 & \Best{.948} & .7376 & .8279\\
\midrule
\cellcolor{orange!18}\textbf{ConTra-0.8B} & .127 & .160 & .514 & .691 & \Second{.936} & \Second{.947} & .507 & .671 & .726 & \Second{.833} & .694 & .795 & .772 & .843 & .765 & .936 & .6302 & .7346\\
\cellcolor{orange!18}\textbf{ConTra-2B} & .988 & \Best{1.000} & .518 & .694 & .922 & .942 & .515 & \Best{.730} & .649 & .825 & .766 & .841 & .785 & .851 & .832 & .937 & .7469 & \Best{.8525}\\
\cellcolor{orange!18}\textbf{ConTra-4B} & \Best{1.000} & \Best{1.000} & .527 & .688 & \Best{.938} & \Best{.952} & .523 & \Second{.692} & .719 & \Best{.840} & .705 & .821 & .754 & .842 & .792 & \Second{.938} & .7448 & \Second{.8465}\\
\bottomrule\end{tabular*}\endgroup
\end{table*}

\section{Experimental Evaluation}
\label{sec:experiments}
We evaluate \textsc{ConTra}'s retrieval quality on ConTEB and analyze the effects of reward components, retrieval model size, and candidate pool size.

\subsection{Experimental Setup}
\label{sec:experiment-setup}
\noindent\textbf{Benchmark.}
We evaluate our method on ConTEB~\cite{DBLP:conf/emnlp/ContiFVBHC25} because it is specifically designed to measure whether retrieval systems can leverage document-level context to disambiguate isolated chunks, and it also serves as a principal benchmark for contextual encoders such as PPLX~\cite{DBLP:conf/acl/EslamiGKMWB26}. Across its eight tasks, ConTEB combines broad domain coverage with distinct forms of contextual dependence. Specifically, MLDR and SQuAD cover encyclopedic text~\cite{DBLP:conf/emnlp/RajpurkarZLL16}, NarrativeQA spans stories and screenplays~\cite{DBLP:journals/tacl/KociskySBDHMG18}, COVID-QA tests biomedical articles~\cite{moller-etal-2020-covid}, and ESG Reports draws from corporate sustainability disclosures, alongside Football, Geography, and Insurance.

\noindent\textbf{Baselines and retrieval components.}
To establish retrieval quality without document context, we first include the general-purpose Qwen3-Embedding-4B model~\cite{DBLP:journals/corr/abs-2506-05176}, which encodes each original chunk independently. We then compare against two categories of contextual retrieval methods: contextual encoders that directly embed surrounding document context into chunk representations, and LLM-based methods that generate explicit textual prefixes prior to standard encoding. Within the contextual-encoder category, we evaluate ConTEB's ModernBERT-Large InSeNT checkpoint trained with in-document negatives using late chunking~\cite{DBLP:conf/emnlp/ContiFVBHC25}, Perplexity's web-scale PPLX-context models at 0.6B and 4B scales where the 4B model sets the prior reported ConTEB state of the art~\cite{DBLP:conf/acl/EslamiGKMWB26}, and Voyage AI's API-based Voyage-context-3 together with its flagship successor Voyage-context-4~\cite{voyage2025context3,voyage2026context4}.

In the textual-prefix category, we compare \textsc{ConTra} against prompt-based contextual retrieval following Anthropic's method~\cite{anthropic2024contextual}, instantiated with untrained Qwen3.5-0.8B, 2B, 4B, and 9B generators~\cite{qwen2026qwen35}. Both \textsc{ConTra} and these prompted baselines prepend the generated prefix to the original chunk and encode the combined text using the same Qwen3-Embedding-4B model. {To compare all methods with a common reranker}, we rerank each method's top 15 candidates with Qwen3-Reranker-4B, which receives original chunks for the context-free and contextual-encoder baselines and prefix-augmented chunks for \textsc{ConTra} and the prompted baselines.

\subsection{Main Results}
\label{sec:main-results}
\noindent\textbf{\textsc{ConTra}-2B achieves state-of-the-art retrieval quality with smaller model scale.}
As shown in Table~\ref{tab:conteb-main}, \textsc{ConTra}-2B reaches an overall reranked nDCG@10 of 0.8525, outperforming the reranked results of the flagship contextual encoders PPLX-4B at 0.7310 and Voyage-context-4 at 0.7300 by over 12 points. {Even against PPLX-4B’s best result, obtained with embedding retrieval alone, \textsc{ConTra}-2B leads by 3.15 points.} Among untrained prompt baselines, 9B is the only model with sufficient capacity to exceed PPLX-4B's strongest result of 0.8210, reaching 0.8279, yet \textsc{ConTra}-2B surpasses this prompted 9B model by 2.46 points, while \textsc{ConTra}-4B outperforms it at 0.8465. These results demonstrate that targeted RL training enables smaller generators to master context selection and replace much larger models, improving indexing efficiency and reducing computational overhead.

\noindent\textbf{\textsc{ConTra} training generalizes across model sizes.}
With the embedding model and reranker held fixed, Table~\ref{tab:conteb-main} shows that RL training improves {both embedding retrieval and reranking} across all three generator sizes. Over same-size Anthropic prompt baselines, \textsc{ConTra} increases final reranked nDCG@10 by 8.15, 9.40, and 3.55 points at 0.8B, 2B, and 4B, respectively.

\noindent\textbf{Model scale shapes generation behavior and task performance.}
Inspecting generated prefixes reveals a clear behavioral divergence between small and large models: smaller generators tend to produce broad, generic topic summaries, whereas larger ones reliably extract specific identifying entities and relations. This contrast appears most clearly on Insurance, which requires matching similarly formatted tables to country headings located elsewhere in a long document. While the 4B and 9B models already reach 1.000 on Insurance prior to training, the 0.8B model performs poorly both before training at 0.199 and after training at 0.160. Thus, RL post-training cannot bridge this gap at 0.8B, as its small parameter capacity limits its ability to parse long-range document structure.

\noindent\textbf{\textsc{ConTra} uses pretrained LLM knowledge without extra pretraining.}
Comparing model families in Table~\ref{tab:conteb-main} underscores the importance of pretrained LLM knowledge for contextual retrieval: InSeNT, built on an encoder-only ModernBERT-Large backbone, reaches only 0.6381 final nDCG@10 and trails every LLM-based method. However, unlocking an LLM's world knowledge and reading comprehension for retrieval does not require architectural modification or pretraining. Whereas PPLX relies on costly diffusion pretraining to convert a causal LLM into a bidirectional encoder before retrieval training~\cite{DBLP:conf/acl/EslamiGKMWB26}, \textsc{ConTra} preserves the generator's native autoregressive text interface and requires only a single epoch of RL post-training with frozen retrieval models.

\subsection{Reward Ablation}
\label{sec:reward-ablation}
\begin{table*}[t]
\centering
\caption{Reward ablation results on ConTEB. Emb. and Rer. report embedding and reranked nDCG@10, respectively. ConTra rows use the full reward objective. Green bold and underlining denote the best and second-best scores within each model scale.}
\label{tab:ablation}

\begingroup
\fontsize{9.5}{10.4}\selectfont
\setlength{\tabcolsep}{1.1pt}
\renewcommand{\arraystretch}{1.0}
\begin{tabular*}{\textwidth}{@{\extracolsep{\fill}}l*{16}{c}cc@{}}
\toprule
Method & \multicolumn{2}{c}{\textbf{Ins.}} & \multicolumn{2}{c}{\textbf{COVID}} & \multicolumn{2}{c}{\textbf{Geo.}} & \multicolumn{2}{c}{\textbf{ESG}} & \multicolumn{2}{c}{\textbf{Football}} & \multicolumn{2}{c}{\textbf{SQuAD}} & \multicolumn{2}{c}{\textbf{NQA}} & \multicolumn{2}{c}{\textbf{MLDR}} & \multicolumn{2}{c}{\textbf{Overall}}\\
\cmidrule(lr){2-3}\cmidrule(lr){4-5}\cmidrule(lr){6-7}\cmidrule(lr){8-9}\cmidrule(lr){10-11}\cmidrule(lr){12-13}\cmidrule(lr){14-15}\cmidrule(lr){16-17}\cmidrule(lr){18-19}
 & Emb. & Rer. & Emb. & Rer. & Emb. & Rer. & Emb. & Rer. & Emb. & Rer. & Emb. & Rer. & Emb. & Rer. & Emb. & Rer. & Emb. & Rer.\\\midrule
\textbf{ConTra-0.8B} & .127 & .160 & \Second{.514} & \Second{.691} & .936 & .947 & .507 & \Best{.671} & \Second{.726} & \Best{.833} & .694 & .795 & \Best{.772} & \Best{.843} & \Second{.765} & \Second{.936} & .6302 & \Best{.7346}\\
0.8B $-\,R_{\mathrm{rank}}$ & \Second{.170} & .161 & .488 & .671 & \Second{.941} & \Best{.957} & \Best{.538} & \Second{.652} & .658 & \Second{.827} & \Best{.735} & \Best{.812} & .748 & .829 & \Best{.805} & \Best{.940} & \Second{.6353} & \Second{.7311}\\
0.8B $-\,R_{\mathrm{fact}}$ & \Best{.202} & \Best{.223} & .488 & .656 & .939 & .953 & \Second{.534} & .595 & .709 & .817 & \Second{.710} & \Second{.809} & \Second{.770} & \Second{.842} & .760 & .911 & \Best{.6388} & .7257\\
0.8B $-\,R_{\mathrm{len}}$ & .152 & \Second{.178} & \Best{.525} & \Best{.692} & \Best{.945} & \Second{.956} & .490 & .637 & \Best{.734} & .820 & .673 & .646 & .766 & .838 & .694 & .806 & .6225 & .6966\\
\midrule
\textbf{ConTra-2B} & \Best{.988} & \Best{1.000} & .518 & \Second{.694} & .922 & .942 & \Best{.515} & \Best{.730} & .649 & \Second{.825} & \Best{.766} & \Best{.841} & \Best{.785} & \Best{.851} & \Best{.832} & \Best{.937} & \Best{.7469} & \Best{.8525}\\
2B $-\,R_{\mathrm{rank}}$ & .140 & \Second{.159} & .520 & .688 & \Best{.941} & \Second{.953} & .508 & .654 & \Second{.702} & \Best{.831} & .699 & \Second{.805} & \Second{.766} & \Second{.840} & \Second{.786} & \Second{.929} & .6326 & .7324\\
2B $-\,R_{\mathrm{fact}}$ & \Second{.985} & \Best{1.000} & \Best{.554} & \Best{.700} & \Second{.929} & \Best{.955} & .469 & .634 & .623 & .813 & .711 & .795 & .737 & .827 & .755 & .913 & \Second{.7202} & \Second{.8297}\\
2B $-\,R_{\mathrm{len}}$ & .136 & .155 & \Second{.528} & .680 & \Best{.941} & .950 & \Second{.510} & \Second{.661} & \Best{.731} & .810 & \Second{.720} & .700 & .756 & .825 & .713 & .840 & .6293 & .7027\\
\midrule
\textbf{ConTra-4B} & \Best{1.000} & \Best{1.000} & .527 & .688 & \Second{.938} & \Second{.952} & .523 & \Best{.692} & \Second{.719} & \Best{.840} & .705 & \Second{.821} & \Second{.754} & \Second{.842} & \Second{.792} & \Best{.938} & .7448 & \Best{.8465}\\
4B $-\,R_{\mathrm{rank}}$ & .985 & \Best{1.000} & \Second{.537} & \Second{.697} & .932 & .949 & \Second{.531} & .661 & .700 & \Second{.828} & \Best{.765} & \Best{.839} & \Best{.772} & \Best{.845} & \Best{.806} & \Second{.932} & \Second{.7535} & \Second{.8440}\\
4B $-\,R_{\mathrm{fact}}$ & \Second{.997} & \Best{1.000} & \Best{.558} & \Best{.705} & \Best{.941} & \Best{.956} & \Best{.572} & \Second{.662} & \Best{.731} & \Best{.840} & \Second{.733} & .803 & .750 & .833 & .768 & .914 & \Best{.7561} & .8391\\
4B $-\,R_{\mathrm{len}}$ & .766 & \Second{.899} & .462 & .598 & .897 & .869 & .424 & .621 & .664 & .608 & .619 & .641 & .727 & .784 & .620 & .648 & .6474 & .7084\\
\bottomrule\end{tabular*}\endgroup
\end{table*}

We remove reranking feedback, factual verification, and length control individually in Table~\ref{tab:ablation}, examining how each component shapes generated prefixes and final retrieval accuracy.

\noindent\textbf{Reranking feedback encourages specific details over generic summaries.}
Tasks such as Insurance test whether a generator can supply concrete identifying details rather than defaulting to broad topic summaries. Because the reranker jointly evaluates the query and augmented chunk, $R_{\mathrm{rank}}$ provides the fine-grained preference signal needed to retain exact attributes. Removing $R_{\mathrm{rank}}$ hits the 2B model hardest, dropping its Insurance nDCG@10 from 1.000 to 0.159, whereas 0.8B and 4B show smaller changes for opposite reasons. The 0.8B model lacks the capacity to parse long-range document structure and stays near 0.16 on Insurance regardless, while the 4B model already extracts country attributes without $R_{\mathrm{rank}}$ and maintains 1.000. Sitting between them, the 2B model can locate country headings in the source document, and $R_{\mathrm{rank}}$ teaches it to keep those identifiers.

\noindent\textbf{Factual verification ensures prefix reliability for retrieval and downstream use.}
At 2B, removing $R_{\mathrm{fact}}$ causes a smaller drop in final retrieval scores than the other ablations, yet it markedly increases factual hallucinations. In a paired manual audit of prefixes for 200 matched target chunks, the factual error rate rises from 4.5\% to 16.0\% when $R_{\mathrm{fact}}$ is omitted. Because nDCG@10 scores annotated chunk identifiers rather than prefix text, fabricated prefixes that match query keywords can still rank the target chunk near the top and even inflate {embedding retrieval scores} at 0.8B and 4B while degrading final reranked nDCG@10. More importantly, in downstream RAG and question-answering pipelines where prefix-augmented chunks serve directly as evidence, unfaithful prefixes corrupt the retrieved context. Adding $R_{\mathrm{fact}}$ suppresses unsupported claims and keeps prefixes reliable across the pipeline.

\noindent\textbf{Length control prevents inter-chunk competition and reduces compute overhead.}
Based on our observations, concise prefixes of roughly 16 tokens suffice to convey essential identifying cues, making lengthy descriptions unnecessary. Removing $R_{\mathrm{len}}$ causes the largest overall retrieval decline at every model size: at 4B, mean prefix length grows from 17.7 to 239.2 tokens while final nDCG@10 falls from 0.8465 to 0.7084. An MLDR document about a Thai wildlife sanctuary illustrates how verbose prefixes induce inter-chunk competition: the full 4B model writes distinct prefixes highlighting each passage's specific topic, whereas removing length control produces lengthy prefixes that repeat the sanctuary's shared location and protected status. Prepending identical background to multiple chunks makes neighboring passages overly similar and causes non-target chunks to match the same query terms. Enforcing brevity avoids this confusion while reducing autoregressive decoding steps and token overhead during indexing and reranking.

\subsection{Generalization Across Retrieval Configurations}
\label{sec:scorer-sensitivity}
{To test whether learned prefixes generalize across retrieval configurations, we keep the prefix generators fixed and vary Qwen3 model size and candidate pool size on ConTEB (Table~\ref{tab:scorer-sensitivity}). For raw Qwen3, Anthropic, and \textsc{ConTra}, we vary both the embedding model and the reranker. InSeNT, PPLX, and Voyage retain their native retrieval embeddings, so only the Qwen3 reranker changes. {{Across all configurations, we also vary the number of retrieved chunks passed to the reranker.}}}

\begin{table}[!htbp]
\centering
\caption{Final nDCG@10 on ConTEB across Qwen3 retrieval model sizes and candidate pool sizes.}
\label{tab:scorer-sensitivity}
\begingroup
\fontsize{9.5}{10.4}\selectfont
\setlength{\tabcolsep}{1.0pt}
\renewcommand{\arraystretch}{1.0}
\begin{tabular*}{\columnwidth}{@{\extracolsep{\fill}}l*{6}{c}@{}}
\toprule
 & \multicolumn{3}{c}{Qwen3 0.6B} & \multicolumn{3}{c}{Qwen3 8B} \\
\cmidrule(lr){2-4}\cmidrule(lr){5-7}
Method & $K{=}10$ & $K{=}15$ & $K{=}20$ & $K{=}10$ & $K{=}15$ & $K{=}20$\\
\midrule
Raw Qwen3 & .5668 & .5792 & .5880 & .6609 & .6735 & .6777\\
InSeNT & .5819 & .6053 & .6065 & .6237 & .6465 & .6526\\
PPLX-0.6B & .6810 & .6649 & .6477 & .7366 & .7295 & .7204\\
PPLX-4B & .7158 & .6862 & .6661 & .7675 & .7454 & .7299\\
Voyage-3 & .7118 & .6839 & .6648 & .7635 & .7445 & .7345\\
Voyage-4 & .7063 & .6856 & .6653 & .7627 & .7467 & .7315\\
\midrule
Anthropic-0.8B & .5763 & .5905 & .5988 & .6525 & .6663 & .6788\\
ConTra-0.8B & .6714 & .6841 & .6898 & .7285 & .7447 & .7513\\
Anthropic-2B & .6966 & .7114 & .7189 & .7689 & .7801 & .7856\\
ConTra-2B & \textbf{.7912} & \textbf{.8045} & \textbf{.8091} & \textbf{.8463} & \textbf{.8558} & \textbf{.8610}\\
Anthropic-4B & .7691 & .7820 & .7905 & .7992 & .8129 & .8239\\
ConTra-4B & \underline{.7808} & \underline{.7978} & \underline{.8068} & \underline{.8397} & \underline{.8541} & \underline{.8568}\\
Anthropic-9B & .7675 & .7794 & .7892 & .8118 & .8219 & .8277\\
\bottomrule
\end{tabular*}
\endgroup
\end{table}

\noindent\textbf{{Learned prefixes transfer across retrieval component scales.}}
{Although trained exclusively with feedback from 4B retrieval models, \textsc{ConTra} generates prefixes that can be used directly with retrieval components of different sizes. At each fixed candidate depth in Table~\ref{tab:scorer-sensitivity}, all three \textsc{ConTra} writers outperform their same-size Anthropic baselines with both the 0.6B and the 8B embedding and reranking models. {At a fixed depth of 15 candidates, for instance, \textsc{ConTra}-2B outperforms Anthropic-2B by 9.31 nDCG@10 points with the 0.6B components and by 7.57 points with the 8B components.}}

\noindent\textbf{{{{Learned prefixes remain effective across candidate pool sizes.}}}}
{Holding each embedding model and reranker fixed, we vary the number of candidates presented to the reranker. All \textsc{ConTra} writers outperform their corresponding prompted baselines at every tested depth, showing that the training gains do not depend on a particular cutoff. With 8B retrieval components, \textsc{ConTra}-2B improves from 0.8463 at ten candidates to 0.8610 at twenty. The 4B components reach 0.8525 with fifteen candidates, within 0.85 points of the peak while using smaller retrieval models and scoring fewer passages per query. To balance retrieval quality and efficiency, we select the 4B pair and fifteen candidates as our default {configuration}.}

\section[RecoBench: Retrieval in Natural and Perturbed Contexts Benchmark]{{{RecoBench}: {Retrieval in Natural and Perturbed Contexts Benchmark}}}
\label{sec:real-data}
While ConTEB specifically targets scenarios where document context is essential to disambiguate or answer a query, real-world retrieval corpora present a more complex landscape. First, practical collections naturally mix self-contained passages---which already contain complete answers independently---with context-dependent chunks that rely on surrounding text; indiscriminate contextualization risks diluting the crisp representation of self-contained chunks. Second, surrounding text is not always helpful: neighboring passages in long documents frequently introduce competing entities, topic shifts, or topically similar distractions that can pull representations away from the target and induce false matches. Existing benchmarks rarely evaluate whether contextual retrieval preserves precision on self-contained passages or resists interference from noisy surroundings. To bridge this gap, we introduce {RecoBench}, combining five realistic application corpora with controlled distraction conditions on NanoBEIR to assess both practical utility and robustness against contextual interference.

\subsection{Dataset Construction}
\label{sec:real-construction}
We cover five application domains: web search, technical question answering, story comprehension, meetings, and financial analysis. For each collection, we retain original questions, split documents into non-overlapping chunks at natural boundaries, and map human-annotated answer spans or supporting evidence to gold chunks. This construction reflects a feature of real collections: some passages can be understood on their own, while others need information from elsewhere in the document. Following ConTEB, we set the chunk size to 1,000 characters, using 2,000 characters for longer Natural Questions pages. Each collection is indexed and searched as a single corpus, with statistics reported in Table~\ref{tab:real-data-stats}.

\begin{table}[!htbp]
\centering
\caption{Sizes of the application collections in {RecoBench}.}
\label{tab:real-data-stats}
\begingroup\fontsize{9.5}{10.4}\selectfont
\setlength{\tabcolsep}{2.5pt}\renewcommand{\arraystretch}{1.0}
\begin{tabular*}{\columnwidth}{@{\extracolsep{\fill}}lccc@{}}
\toprule
\multicolumn{1}{c}{Collection} & Queries & Docs. & Chunks \\
\midrule
NQ & 5,499 & 5,181 & 99,096 \\
TechQA & 149 & 139 & 1,036 \\
FairytaleQA & 1,007 & 23 & 342 \\
QMSum & 244 & 35 & 2,140 \\
FinanceBench & 146 & 84 & 45,049 \\
\bottomrule
\end{tabular*}\endgroup
\end{table}

\noindent\textbf{Natural Questions (NQ).}
NQ comprises real Google search queries paired with human-annotated answers on Wikipedia pages~\cite{DBLP:journals/tacl/KwiatkowskiPRCP19}. We preserve page order and map annotated answer spans to chunks, using short-answer spans to pinpoint evidence whenever a long answer spans multiple chunks.

\noindent\textbf{TechQA.}
TechQA features technical support questions from IBM forums grounded in IBM Technotes~\cite{DBLP:conf/acl/CastelliCDFFFGK20}. Using its answerable development queries, we split source documents into chunks and designate the chunk with the highest overlap with the annotated answer span as gold.

\noindent\textbf{FairytaleQA.}
{FairytaleQA~\cite{DBLP:conf/acl/XuWYRYWZLBSHS0M22} pairs expert-written questions with supporting sections, which we map to chunks in story order to retain a natural mix of local and cross-section queries. {{ConTEB's NarrativeQA~\cite{DBLP:journals/tacl/KociskySBDHMG18} also uses stories, but its design encourages questions that draw on different parts of the narrative. This makes it well suited to testing when broader context helps, while our FairytaleQA collection preserves the varying context needs of the original task.}}}

\noindent\textbf{QMSum.}
QMSum pairs meeting transcripts with human-written queries and annotated supporting utterances for summarization~\cite{DBLP:conf/naacl/ZhongYYZMJACLQR21}. We use its specific queries as retrieval questions and construct gold chunks from supporting utterances while preserving speaker labels and turn order.

\noindent\textbf{FinanceBench.}
FinanceBench contains specialist-curated financial questions, answers, and supporting evidence from corporate filings~\cite{DBLP:journals/corr/abs-2311-11944}. We preserve readable table formatting during chunking and select the chunk with the greatest overlap with the annotated evidence as gold, supporting queries over company names, reporting periods, and numerical figures.

\begin{table*}[!t]
\centering
\caption{{RecoBench} nDCG@10. FQA and Finance denote FairytaleQA and FinanceBench; Orig., Rand., and Topic. denote original, random, and topical context conditions.}
\label{tab:real-detailed}

\begingroup
\fontsize{9.5}{10.4}\selectfont
\setlength{\tabcolsep}{1.1pt}
\renewcommand{\arraystretch}{1.13}
\begin{tabular*}{\textwidth}{@{\extracolsep{\fill}}l*{16}{c}cc@{}}
\toprule
 & \multicolumn{10}{c}{\cellcolor{gray!12}Application collections} & \multicolumn{6}{c}{\cellcolor{gray!12}NanoBEIR context} & \multicolumn{2}{c}{}\\
\cmidrule(lr){2-11}\cmidrule(lr){12-17}
Method & \multicolumn{2}{c}{\textbf{NQ}} & \multicolumn{2}{c}{\textbf{TechQA}} & \multicolumn{2}{c}{\textbf{FQA}} & \multicolumn{2}{c}{\textbf{QMSum}} & \multicolumn{2}{c}{\textbf{Finance}} & \multicolumn{2}{c}{\textbf{Orig.}} & \multicolumn{2}{c}{\textbf{Rand.}} & \multicolumn{2}{c}{\textbf{Topic.}} & \multicolumn{2}{c}{\textbf{Overall}}\\
\cmidrule(lr){2-3}\cmidrule(lr){4-5}\cmidrule(lr){6-7}\cmidrule(lr){8-9}\cmidrule(lr){10-11}\cmidrule(lr){12-13}\cmidrule(lr){14-15}\cmidrule(lr){16-17}\cmidrule(lr){18-19}
 & Emb. & Rer. & Emb. & Rer. & Emb. & Rer. & Emb. & Rer. & Emb. & Rer. & Emb. & Rer. & Emb. & Rer. & Emb. & Rer. & Emb. & Rer.\\\midrule
Qwen3-Emb.-4B & .651 & .710 & .722 & .742 & .628 & .787 & .314 & .357 & .292 & .339 & \Best{.689} & \Best{.721} & \Best{.689} & \Best{.721} & \Best{.689} & \Best{.721} & .5845 & .6373\\
InSeNT & .602 & .686 & .653 & .738 & .494 & .720 & .110 & .192 & .088 & .104 & .516 & .606 & .398 & .551 & .494 & .635 & .4194 & .5291\\
PPLX-0.6B & .598 & .685 & .728 & .758 & \Second{.674} & .798 & .263 & .301 & .247 & .288 & .597 & .685 & .582 & .668 & .596 & .683 & .5357 & .6083\\
PPLX-4B & .679 & .717 & .759 & .770 & \Best{.722} & \Second{.812} & \Best{.412} & .405 & .418 & .445 & .629 & .705 & .590 & .678 & .633 & .703 & .6051 & .6542\\
Voyage-3 & \Second{.691} & .735 & \Second{.820} & .779 & .657 & .777 & .377 & \Second{.410} & \Second{.522} & .479 & .651 & .710 & .652 & \Second{.713} & .642 & .706 & \Second{.6265} & .6636\\
Voyage-4 & \Best{.719} & \Second{.741} & \Best{.827} & .779 & .651 & .772 & \Second{.379} & .402 & \Best{.524} & .469 & .677 & .716 & .623 & .698 & .652 & .709 & \Best{.6315} & .6608\\
\midrule
Qwen3.5-0.8B & .651 & .705 & .687 & .752 & .611 & .778 & .320 & .349 & .325 & .415 & .667 & .705 & .666 & .703 & .659 & .700 & .5733 & .6384\\
Qwen3.5-2B & .672 & .713 & .733 & .755 & .632 & .799 & .323 & .372 & .333 & .443 & .661 & .703 & .647 & .691 & .652 & .690 & .5816 & .6456\\
Qwen3.5-4B & .688 & .731 & .710 & \Second{.788} & .640 & .811 & .339 & .388 & .331 & .475 & .663 & .698 & .617 & .661 & .646 & .691 & .5793 & .6553\\
Qwen3.5-9B & .681 & .727 & .735 & \Best{.795} & .631 & .798 & .371 & .404 & .369 & \Second{.494} & .657 & .703 & .641 & .692 & .650 & .693 & .5918 & .6632\\
\midrule
\cellcolor{orange!18}\textbf{ConTra-0.8B} & .656 & .725 & .706 & .771 & .636 & .796 & .299 & .347 & .334 & .412 & \Second{.686} & .718 & .667 & .698 & \Second{.676} & .711 & .5825 & .6471\\
\cellcolor{orange!18}\textbf{ConTra-2B} & .683 & \Best{.749} & .692 & .765 & .661 & .809 & .338 & .381 & .402 & \Best{.525} & .679 & .709 & \Second{.678} & .710 & .673 & .706 & .6008 & \Best{.6692}\\
\cellcolor{orange!18}\textbf{ConTra-4B} & .672 & .727 & .711 & .752 & .643 & \Best{.812} & .358 & \Best{.411} & .391 & .491 & .680 & \Second{.719} & .669 & .707 & .671 & \Second{.715} & .5992 & \Second{.6669}\\
\bottomrule\end{tabular*}\endgroup
\end{table*}

\noindent\textbf{NanoBEIR context conditions.}
NanoBEIR is a small-scale benchmark of 13 BEIR tasks spanning domains such as news, scientific literature, and question answering~\cite{DBLP:conf/nips/Thakur0RSG21,zetaalpha_nanobeir}. Because standard BEIR passages are largely self-contained, NanoBEIR provides an ideal testbed to operationalize the two failure modes motivated above: whether contextualization dilutes self-contained passages or allows noisy neighbors to corrupt target representations. To isolate the effect of surrounding text from corpus differences, we hold queries, target chunks, and gold annotations strictly fixed while varying only surrounding context across three controlled conditions: the \emph{original} condition presents each target chunk alone as a clean baseline; the \emph{random} condition prepends two randomly sampled passages before it and appends two after it; and in the \emph{topical} condition, GPT-6 Luna~\cite{openai2026luna} generates four independent passages on the target's narrow topic, each describing an unrelated subject, case, or event. Every contextual method receives identical surrounding passages. Random neighbors simulate abrupt topic shifts, while topical neighbors simulate passages that discuss the same subject area but provide no evidence about the target case.

\subsection{{RecoBench} Results}
\label{sec:real-results}
We evaluate \textsc{ConTra} on {RecoBench} using the embedding model, reranker, and top-15 retrieval pipeline from Section~\ref{sec:experiment-setup}, reporting nDCG@10 across the five application collections and three NanoBEIR conditions in Table~\ref{tab:real-detailed}.

\noindent\textbf{\textsc{ConTra} generalizes to new application tasks.}
{{{{The generator \authorrevision{receives RL training} only on encyclopedic and narrative text, while {RecoBench} extends evaluation to \authorrevision{domains unseen during this training, such as} finance, technical support, and meetings.}} }Without further training on these collections, \textsc{ConTra}-2B still outperforms its same-size prompted baseline on all five. Its overall reranked nDCG@10 across the eight settings rises from 0.6456 to 0.6692. The application queries retain their original mix of context requirements. Gains on these collections show that the learned prefix policy transfers both to new domains and to settings in which the value of neighboring passages varies from query to query.}

\noindent\textbf{Different model sizes offer complementary strengths across tasks.}
RL training improves performance across most settings at all three scales, with each size excelling on distinct task types. \textsc{ConTra}-2B achieves the strongest overall balance and broadest gains across the benchmark. \textsc{ConTra}-4B leads all trained variants on QMSum and FairytaleQA, where greater capacity helps track speaker turns in meetings or resolve character aliases across long narratives. Conversely, TechQA exhibits a reversed size ordering within the trained family, where \textsc{ConTra}-0.8B at 0.7713 outperforms \textsc{ConTra}-4B at 0.7522. Because technical support passages frequently already contain the target error message or procedure, the smaller model's broader topical summary establishes relevance without over-narrowing the chunk to a single sub-procedure when a query targets another aspect of the same passage.

\begin{table}[!htbp]
\centering
\caption{Comparison of candidate retrieval using ConTra versus PPLX-4B, {with ConTra prefixes supplied to the reranker in both cases}. Reported metrics are reranked nDCG@10. Bold marks the stronger retriever within each generator size.}
\label{tab:hybrid}
\begingroup\fontsize{9.5}{10.4}\selectfont
\setlength{\tabcolsep}{2.5pt}\renewcommand{\arraystretch}{1.0}
\begin{tabular*}{\columnwidth}{@{\extracolsep{\fill}}lcccc@{}}
\toprule
 & \multicolumn{2}{c}{2B model} & \multicolumn{2}{c}{4B model} \\
\cmidrule(lr){2-3}\cmidrule(lr){4-5}
\multicolumn{1}{c}{Collection} & ConTra & PPLX & ConTra & PPLX \\
\midrule
\multicolumn{5}{l}{\textit{Application collections}} \\
NQ & \textbf{.7486} & .7382 & \textbf{.7267} & .7161 \\
TechQA & .7649 & \textbf{.7797} & .7522 & \textbf{.7834} \\
FairytaleQA & .8092 & \textbf{.8194} & .8123 & \textbf{.8234} \\
QMSum & .3811 & \textbf{.4140} & .4110 & \textbf{.4286} \\
FinanceBench & .5248 & \textbf{.5250} & .4913 & \textbf{.5291} \\
\midrule
\multicolumn{5}{l}{\textit{NanoBEIR conditions}} \\
Original & \textbf{.7088} & .6993 & \textbf{.7190} & .7016 \\
Random & \textbf{.7102} & .6729 & \textbf{.7071} & .6718 \\
Topical & \textbf{.7057} & .6909 & \textbf{.7154} & .7030 \\
\midrule
\textbf{Overall} & \textbf{.6692} & .6675 & .6669 & \textbf{.6696} \\
\bottomrule
\end{tabular*}\endgroup
\end{table}

\noindent\textbf{{{{Reranking generally helps on real-world collections.}}}}
{Reranking raw chunks lowers PPLX-4B accuracy on ConTEB. On the five {RecoBench} application collections, which retain natural queries, the same reranker improves PPLX-4B on four. The benefit extends across methods: reranking improves nDCG@10 in 60 of the 65 method--collection pairs. \authorrevision{These results indicate that multi-stage retrieval is more meaningful in real-world scenarios, and evaluating a model solely on the results of a single stage provides an incomplete assessment.} \textsc{ConTra}-2B \authorrevision{supports both stages with} a shared readable prefix and gains from reranking on all five collections.}

\noindent\textbf{Training strengthens robustness against distracting context.}
PPLX encodes surrounding passages into the target's vector, making retrieval susceptible to distracting neighbors. The prefix-based route instead preserves the original chunk verbatim and adds context as a prefix, retaining the target's own content as a stable anchor. Even before RL, random context reduces reranked nDCG@10 by only 1.19 points for the prompted 2B prefix baseline, compared with 2.66 points for PPLX-4B. Our training further strengthens this robustness. On NanoBEIR, \textsc{ConTra}-4B raises mean reranked nDCG@10 across the three conditions from 0.6831 to 0.7138 after training, with the largest gain under random context, while \textsc{ConTra}-2B also improves under both random and topical interference. With both the prefix architecture and model size held fixed, these additional gains reflect learned generation behavior. Although training data is filtered for context dependence, some surrounding passages offer little useful information or are difficult to interpret reliably. Relevance rewards and factual verification encourage the generator to distill and reinforce the target's own content in these cases. This learned selectivity further reduces the influence of random or topically similar distractors, strengthening the protection provided by preserving the original chunk.

\subsection{Combining \textsc{ConTra} Prefixes with PPLX Retrieval}
\label{sec:hybrid}
Although standalone \textsc{ConTra} achieves strong end-to-end retrieval quality, its text-based output interface allows direct pairing with existing contextual encoders. Extending our observation in Section~\ref{sec:observation} that reranker-visible prefixes unlock PPLX-4B's strong candidate recall, we test a hybrid pipeline that uses PPLX-4B {to retrieve candidates} and attaches \textsc{ConTra} prefixes {before reranking them}, holding the generator and reranker fixed in Table~\ref{tab:hybrid}. Overall averages are closely matched at both 2B and 4B, yet task-level results show distinct performance patterns across settings.

\noindent\textbf{\textsc{ConTra} prefixes {improve the reranking of candidates from contextual encoders}.}
With PPLX-4B candidates held fixed, attaching \textsc{ConTra}-2B prefixes improves reranked nDCG@10 across all five application collections compared with reranking raw chunks---most notably on FinanceBench, where nDCG@10 rises from 0.4446 to 0.5250. PPLX utilizes document-aware embeddings to recall relevant passages, while \textsc{ConTra} externalizes key attributes such as company names and reporting periods into explicit text that the reranker can verify.

\begin{figure*}[t]
\centering
\includegraphics[width=0.94\textwidth]{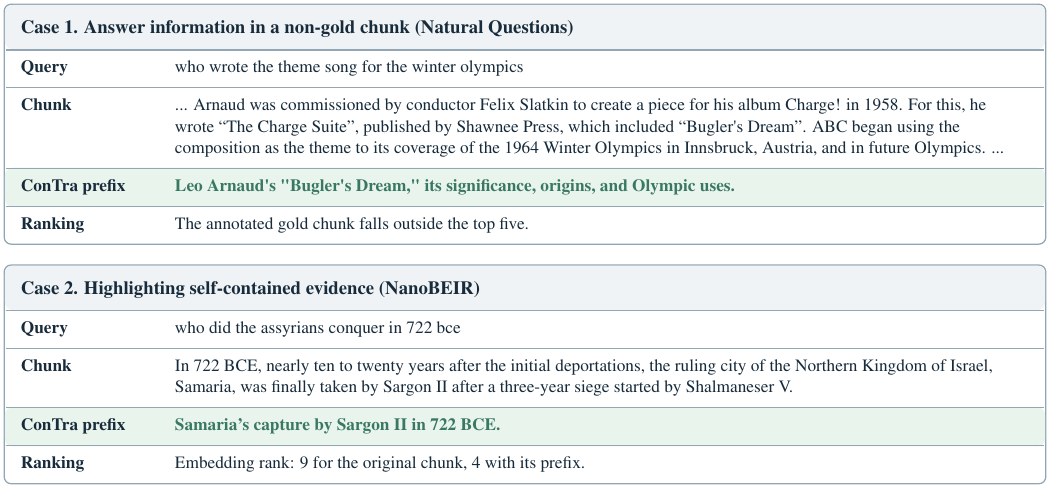}
\caption{Case 1: A prefix supplies answer information to a non-gold chunk. Case 2: A prefix highlights relevant information already present in the target chunk.}
\Description{Two vertically stacked bordered cases. The first shows Leo Arnaud's full name added to a neighboring passage. The second shows a passage about the capture of Samaria moving from ninth to fourth in embedding retrieval when its prefix is included.}
\label{fig:context-cases}
\end{figure*}

\noindent\textbf{{Better candidates from PPLX} improve hybrid retrieval.}
Comparing individual collections at 2B shows that the hybrid configuration outperforms standalone \textsc{ConTra} on QMSum, FairytaleQA, and TechQA, lifting QMSum from 0.3811 to 0.4140. On meeting and story tasks, PPLX's document-level representations help retrieve passages whose relevance depends on speaker or narrative relationships, while TechQA gains confirm that a strong candidate retriever complements the same readable prefixes across distinct task needs. Because a reranker can only reorder candidates {returned by the retriever}, combining PPLX's broader contextual recall with \textsc{ConTra}'s explicit prefixes {improves accuracy} on these collections.

\noindent\textbf{{{{Standalone \textsc{ConTra} is more robust when surrounding context is uninformative.}}}}
{With the prefixes and reranker held fixed, standalone \textsc{ConTra} outperforms the hybrid across all three NanoBEIR conditions at both generator sizes. Under random interference, for example, the 2B model scores 0.7102 versus 0.6729. {The gap between the two configurations is largest under random interference at both model sizes. These results favor standalone \textsc{ConTra} {when \authorrevision{neighboring passages do not provide useful context}.}}}

\section{Cost and Efficiency}
\label{sec:efficiency}

Practical deployment of contextual retrieval requires balancing retrieval accuracy against indexing latency and storage overhead. In this section, we evaluate the computational and storage costs of \textsc{ConTra} across model scales and compare them against standard dense retrieval, dedicated contextual encoders, and untrained~prefix~generators.

\subsection{Measurement Protocol}
\label{sec:measurement_protocol}

We benchmark indexing latency on a single NVIDIA H100 80GB GPU across two subsets chosen to represent distinct document length regimes: 51 long Wikipedia articles from Natural Questions comprising 1,027 chunks, and 23 short narratives from FairytaleQA comprising 342 chunks. All methods process identical chunks under the same source-context budget. After warming up each model, we execute three timed runs per configuration and report median end-to-end latency, covering prefix generation where applicable and dense embedding computation while excluding one-time initialization overhead. To ensure a fair comparison across architectures, we evaluate PPLX-4B using its official ONNX-optimized implementation and run all Qwen3.5 generators on vLLM.

\begin{table}[!htbp]
\centering
\caption{Total indexing time in seconds on a single NVIDIA H100 80GB GPU. Paired Qwen3.5–ConTra runs hold all inference settings identical within each model size.}
\label{tab:cost}
\label{tab:paired-cost}
\begingroup\fontsize{9.5}{10.4}\selectfont
\setlength{\tabcolsep}{2.5pt}\renewcommand{\arraystretch}{1.0}
\begin{tabular*}{\columnwidth}{@{\extracolsep{\fill}}lcc@{}}
\toprule
\multicolumn{1}{c}{Method} & Long docs. & Short docs. \\
\midrule
Qwen3-Emb.-4B & 7.92 & 1.01 \\
PPLX-0.6B & 29.97 & 3.51 \\
PPLX-4B & 77.97 & 8.08 \\
Qwen3.5-9B & 268.12 & 29.02 \\
\midrule
Qwen3.5-0.8B & 53.39 & 6.20 \\
\textbf{ConTra-0.8B} & 38.09 & 2.70 \\
Qwen3.5-2B & 76.58 & 7.87 \\
\textbf{ConTra-2B} & 55.58 & 3.75 \\
Qwen3.5-4B & 124.30 & 14.05 \\
\textbf{ConTra-4B} & 114.19 & 10.11 \\
\bottomrule
\end{tabular*}\endgroup
\end{table}

\subsection{Indexing Time}
\label{sec:indexing_time}

\noindent\textbf{\textsc{ConTra}-2B substantially reduces indexing latency relative to 9B and PPLX-4B.}
As reported in Table~\ref{tab:cost}, \textsc{ConTra}-2B indexes the short-document collection $7.74\times$ as fast as the prompted 9B pipeline, cutting total latency from 29.02\,s to 3.75\,s, and $2.15\times$ as fast as PPLX-4B at 8.08\,s. On the long-document collection, \textsc{ConTra}-2B maintains a consistent speed advantage, indexing $4.82\times$ as fast as prompted 9B and $1.40\times$ as fast as PPLX-4B. Within the trained \textsc{ConTra} family, \textsc{ConTra}-0.8B achieves the lowest overall latency across both collections, whereas \textsc{ConTra}-2B combines the strongest retrieval accuracy with substantially lower indexing cost than \textsc{ConTra}-4B.

\noindent\textbf{Prefix caching yields substantially higher reuse on short documents.}
{vLLM's prefix cache reuses attention keys and values for the input tokens that requests share at the start of a prompt. Chunks from the same short document share the full source-document prompt, so this context can be processed once across requests. Long documents use different context windows with fewer shared leading tokens. The cache hit rate reaches 81.9\% on short documents versus 24.9\% on long ones, helping explain the larger relative speedup on short documents.}

\noindent\textbf{RL training accelerates inference at fixed model scale.}
Comparing paired runs before and after training---with prompts, input budgets, batch sizes, and vLLM settings held identical within each model size---demonstrates that RL post-training consistently lowers indexing latency across all three generator scales. At 2B, short-document indexing time drops from 7.87\,s to 3.75\,s, driven by a 71.1\% reduction in generated output length from 82.4 to 23.8 tokens per chunk, compared with 79.7 tokens for prompted 9B. Generating concise prefixes saves both autoregressive decoding steps and input tokens passed to the downstream embedding model. Because high source-cache reuse leaves autoregressive decoding as a larger fraction of runtime on short documents, output length compression translates into wall-clock savings, aligning indexing efficiency with the inter-chunk distinctiveness observed in Section~\ref{sec:reward-ablation}.

\subsection{Disk Storage}
\label{sec:disk_storage}

Retaining generated prefixes adds a lightweight text payload to the index without altering its vector layout. Specifically, each prefix-augmented chunk is encoded into a single 2,560-dimensional \texttt{float32} vector, keeping vector count and dimensionality unchanged. Measuring the UTF-8 size of stored prefixes and separators relative to raw chunk text plus dense vectors, the median of three saved runs adds 36.1\,KB on FairytaleQA and 98.2\,KB on Natural Questions. The corresponding original text-and-vector payloads are 3.77\,MB and 12.46\,MB, so the additions represent only 0.96\% and 0.79\%, respectively. \textsc{ConTra} thus retains a standard single-vector representation while adding less than 1\% in text payload to keep readable context available at retrieval time.

\section{Case Studies}
\label{sec:cases}

While aggregate ranking metrics quantify overall retrieval accuracy, inspecting individual predictions reveals how generated prefixes reshape retrieved evidence in practice. Figure~\ref{fig:context-cases} examines two cases from \textsc{ConTra}-2B illustrating complementary behaviors across distinct contextual regimes: recovering missing entity information into an adjacent passage in Natural Questions, and foregrounding existing defining features within an isolated passage in NanoBEIR.

\noindent\textbf{{Case 1. Recovering an answer from surrounding context.}} The first query asks for the Winter Olympics theme composer, whose annotated answer is Leo Arnaud. \textsc{ConTra}-2B ranks an adjacent passage first while the annotated gold chunk falls outside the top five, so the first-ranked passage contributes zero to nDCG@10 under the evaluation used above. However, because that passage already describes the composition's Olympic use and sits adjacent to the gold chunk, \textsc{ConTra}'s prefix extracts the composer's full name from the surrounding page and prepends it directly, supplying the answer at rank 1. This indicates that retrieval-only, chunk-ID-based metrics can underestimate \textsc{ConTra}'s effectiveness, as its prefixes not only aid retrieval but also carry self-contained answer cues that could be reused by downstream tasks such as question answering.

\noindent\textbf{{Case 2. Distilling key facts from an isolated passage.}} {The second case illustrates how the generator can foreground the target’s own content when surrounding context is unavailable (Section~\ref{sec:real-results}).} In this NanoNQ example, the query asks who the Assyrians conquered in 722 BCE, while the isolated passage embeds the answer among broader historical context. Seeing only the chunk itself, \textsc{ConTra}-2B distills the core facts directly from the passage---Samaria's capture by Sargon II in 722 BCE---into a concise prefix, making the passage easier to retrieve. Prepending this prefix raises the gold chunk's embedding retrieval rank from 9 to 4.

Because our relevance rewards measure ranking gains over the unaugmented target chunk, RL training incentivizes the generator not only to extract external context, but also to foreground the target's core facts when surrounding text adds little signal. At inference, this learned distillation transfers to self-contained passages, making their defining information more explicit to both encoder and reranker.

\section{Conclusion}
\label{sec:conclusion}

{Readable prefixes can be reused across embedding retrieval and reranking. Building on this potential, we introduced \textsc{ConTra}, a reinforcement learning framework that trains small prefix generators using embedding and reranking feedback, factual verification, and length control. Trained only on ConTEB, \textsc{ConTra}-2B outperforms PPLX-4B, Voyage-4, and a prompted 9B model using Anthropic's method on both ConTEB and {RecoBench}, achieving the highest overall reranking score on {RecoBench}. On short documents, it indexes $7.74\times$ as fast as the 9B baseline and $2.15\times$ as fast as PPLX-4B. {{{Future work could tailor prefix generator training to specific downstream tasks, such as question answering, so that the generated context serves the objective of each task.}}}}

\bibliographystyle{ACM-Reference-Format}
\bibliography{references,sections/related_references,sections/additional_references,sections/revision_references}
\end{document}